\documentclass[reprint,prd, superscriptaddress, tightenlines, longbibliography, nofootinbib, eqsecnum, amsfonts, amsmath, floatfix, notitlepage, onecolumn]{revtex4-1}
\usepackage{pgfplots}
\usepackage{hyperref}

\usepackage[a4paper, total={7.1in, 9in}]{geometry}
\usepackage{amsmath} 
\usepackage{amssymb}
\usepackage[utf8]{inputenc}
\usepackage{color}

\newcommand{\Cov}[0]{{\textrm{Cov}}}

\newcommand{\si}[0]{{s_{\rm i}}}
\newcommand{\ti}[0]{{t_{\rm i}}}

\newcommand{\so}[0]{{s_{\rm o}}}
\renewcommand{\to}[0]{{t_{\rm o}}}
\newcommand{\uo}[0]{{u_{\rm o}}}
\newcommand{\vo}[0]{{v_{\rm o}}}

\newcommand{\Ylm}[1]{\:_{#1}Y_{\ell m}}

\newcommand{\av}[1]{\left\langle #1 \right\rangle}
\newcommand{\resp}{ {\mathcal R} }

\newcommand{\hn}[0]{\hat n}
\begin{document}
\title{Position-space curved-sky anisotropy quadratic estimation}
\begin{abstract}This document supplements the public release of the Planck 2018 CMB lensing pipeline, made available at~\url{https://github.com/carronj/plancklens}. It collects calculations relevant to curved-sky separable quadratic estimators in the spin-weight, position-space correlation function formalism, including analytic calculations of estimator responses and Gaussian noise biases between arbitrary pairs of quadratic estimators. It also contains the derivation of optimal, joint gradient and curl mode quadratic estimators for parametrized anisotropy of arbitrary spin. 
\end{abstract}	
\author{Julien Carron}

\maketitle
\tableofcontents

\vspace{1cm}
This document supplements the release of the Planck 2018 CMB lensing \cite{Aghanim:2018oex} pipeline, now available at~\url{https://github.com/carronj/plancklens}. It collects calculations relevant to curved-sky separable quadratic estimators in the spin-weight, position-space correlation function formalism, including analytic calculations of estimator responses (Section~\ref{sec:responses}, Eqs.~(\ref{eq:RLgc} - \ref{eq:Rxipsi})) and Gaussian noise biases (Section~\ref{sec:biases}, Eqs.~(\ref{eq:N0gc} - \ref{eq:N0C}) between arbitrary pairs of quadratic estimators. It also contains the derivation of optimal, joint gradient and curl mode quadratic estimators for parametrized anisotropy of arbitrary spin (Section~\ref{sec:opti}). Examples follow in Section~\ref{sec:examples}. In this formalism, minimum variance (MV) estimators combining temperature and polarization can typically be obtained with at most 6 spin-weight harmonic transforms (gradient and curl jointly), and responses and noise-covariances with a series of one-dimensional Wigner small-$d$ transforms, as first proposed in Refs~\citep{Dvorkin:2009ah, Smith:2007rg}
\section{Spin-weight estimators}
A complex spin-$s$ field $_{s}f(\hn)$ on the sphere is defined with reference to local axes by the condition that it transforms under a clockwise rotation of angle $\psi$ of the local axes according to $_{s}f(\hn) \rightarrow e^{i s \psi}\:_{s}f(\hn)$. We use further the notation $_{-s}f(\hn) = \:_{s}f^*(\hn)$.

The gradient (g) and curl (c) harmonic modes of definite parity of $_sf$ are then defined as follows
\begin{eqnarray}\label{eq:gc1}
		g^{s}_{\ell m} &= -\frac 12\left(\:_{|s|} f_{\ell m} + (-1)^s \:_{-|s|} f_{\ell m}\right)\\ \label{eq:gc2}
		c^{s}_{\ell m} &=-\frac 1{2i} \left( \:_{|s|} f_{\ell m} - (-1)^s \:_{-|s|} f_{\ell m} \right)
\end{eqnarray}
where  $_{\pm s} f_{\ell m} \equiv \int d^2n \:_{\pm s}f(\hn) \:_{\pm s}Y^*_{\ell m}(\hn)$. Spin-0 fields are real and pure gradients.
With these conventions, we have in particular for the CMB polarization $_{\pm 2} P(\hn)$
\begin{equation}
	_{\pm 2}P_{\ell m} = -(E_{\ell m} \pm i B_{\ell m}).
\end{equation}
where $E$ and $B$ are the polarization gradient and curl modes\footnote{The spin-0 intensity $_0T(\hn)$ gradient mode is $-T_{\ell m}$ and not $T_{\ell m}$}. Our polarization conventions are such that $_{\pm 2} P(\hn) = Q(\hn) \pm i U(\hn)$, with the local x and y axes at each point $\hn$ pointing south and east (following e.g. the healpix\cite{Gorski:2004by}~software conventions, or those of Ref.~\cite{Lewis:2006fu}, but differing from the IAU standards, see \url{https://healpix.jpl.nasa.gov/html/intronode12.htm}).
\newline
The relation inverse to Eqs.~\eqref{eq:gc1} and~\eqref{eq:gc2} is
\begin{equation}\label{eq:gcinverse}
	_{\pm |s|} f_{\ell m} =- (\pm)^s\left( g^s_{\ell m} \pm i c^s_{\ell m} \right).
\end{equation}
\subsection{Correlation functions}
\begin{figure}[h!]
	\includegraphics[width=0.49\textwidth]{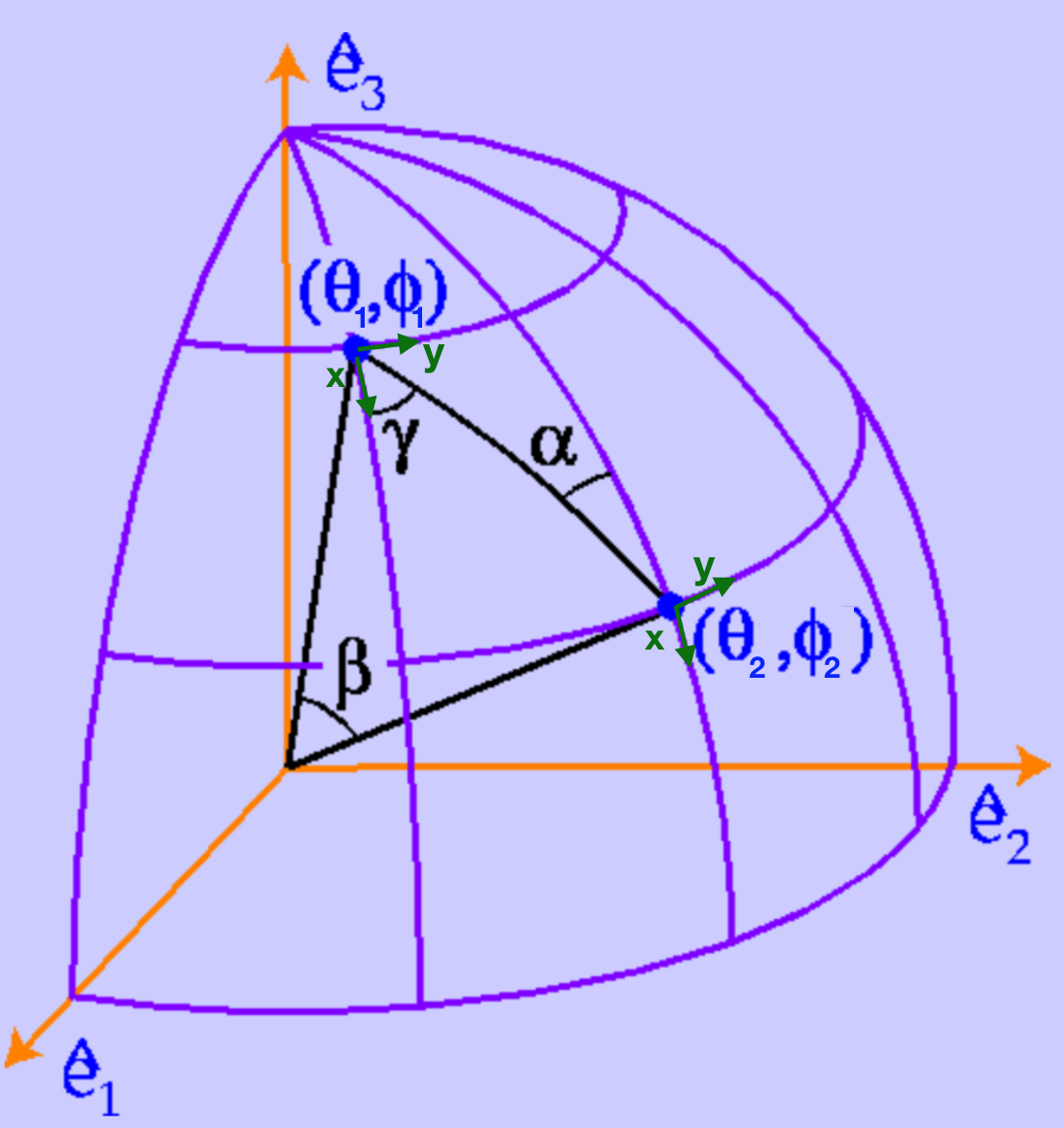}
	\caption{\label{fig:geometry}The geometry and angles in Eq.~\eqref{eq:cf}, with the local axes in green. Figure originally from Wayne Hu tutorials, \url{http://background.uchicago.edu/~whu/tamm/webversion/node5.html}.}
\end{figure}

We use position-space correlation function for fields of arbitrary spins as follows. As is well-known from the case of CMB polarization, in order to undo the dependence on the local axis definition, the fields must first be defined with respect to a common relevant basis~\cite[e.g.]{Chon:2003gx, Challinor:2005jy}. For two points on the sphere $\hn_1$, $\hn_2$, let $\gamma$ be the angle at $\hn_1$ between the local $x$-axis to the geodesic connecting $\hn_1$ and $\hn_2$, and $\alpha$ defined in the same way at $\hn_2$. See Fig.~\ref{fig:geometry} for the geometry. Clockwise rotations by $\pi - \gamma$ at $\hn_1$ and by $\pi - \alpha$ at $\hn_2$ align the local bases, and we may define

\begin{equation}\label{eq:cf}
\begin{split}
	\xi^{st}_{+}(\beta) &\equiv \av{e^{is(\pi-\gamma)}\:_{s}f(\hn_1)\:\left(_{t}f(\hn_2)e^{i t(\pi-\alpha)}\right)^*}\\
		\xi^{st}_{-}(\beta)& \equiv \av{ \left(e^{is(\pi-\gamma)}\:_{s}f(\hn_1)\right)^*\:\left(_{t}f(\hn_2)e^{i t(\pi- \alpha)}\right)^*}.
\end{split}
\end{equation}
In harmonic space, and using relation~\eqref{eq:gcinverse} gives the following expression
\begin{widetext}
\begin{equation}
\begin{split}
	\xi_{+}^{st}(\beta) =\left(+1\right)^s &\sum_{\ell} \left(\frac{2\ell + 1}{4\pi}\right) \left[C_\ell^{g^sg^{t}} + C_\ell^{c^sc^{t} }-i\left(C_\ell^{g^sc^{t}} - C_\ell^{c^sg^{t}}\right)\right]d^\ell_{s t}(\beta) \\
	\xi_{-}^{st}(\beta) = \left(-1\right)^s &\sum_{\ell} \left(\frac{2\ell + 1}{4\pi}\right) \left[C_\ell^{g^sg^{t}} - C_\ell^{c^sc^{t} }-i\left(C_\ell^{g^sc^{t}} + C_\ell^{c^sg^{t}}\right)\right]d^\ell_{-s t}(\beta) 
\end{split}
\end{equation}
\end{widetext}
where $\beta$ is the distance between $\hn_1$ and $\hn_2$.
These two correlation functions carry all of the information on their gradient and curl mode spectra.
\subsection{Quadratic estimators}
Prior to projection onto gradient and curl modes, and prior to proper normalization, separable quadratic estimators can be written as a (sum of) products of two position-space maps. Let $\hat q$ be such an unnormalized estimator: \begin{equation}\label{eq:QE}
\begin{split}	
 _{s_o + t_o}\hat q(\hn) \equiv &\left(\sum_{\ell m}\: w^{\so\si}_\ell \:_{\si} \bar X_{\ell m} \:_\so Y_{\ell m}(\hn)\right)\left(\sum_{\ell m}\:w^{\to\ti}_\ell  \:_{\ti} \bar X_{\ell m} \:_\to Y_{\ell m}(\hn)\right)
 \end{split} 
\end{equation}
where $\si, \ti$ are input spins, $\so, \to$ outputs spins, and $w_\ell^{\so\si}, w_\ell^{\ti\to}$ associated weights. For simplicity we use the same symbol $w$ for the first and second leg weights even if they may differ in general for the same spin indices. By consistency with $_{-s_o - t_o} \hat q(\hn) = \:_{s_o + t_o}\hat q^*(\hn)$ the weights have symmetry $w_\ell^{-\so-\si} = (-1)^{\so + \si}w_\ell^{*\so \si}$.
\newline
\newline
The maps $_s \bar X_{lm}$ are the inverse signal + noise variance filtered CMB maps; the filtered scalar temperature
\begin{equation}
	_0 \bar X_{\ell m} = \bar T_{\ell m}
\end{equation}	
and filtered spin $\pm 2$ Stokes polarization $_{\pm 2}\bar P = \bar Q \pm i\bar U$,
\begin{equation}
\quad _{\pm 2} \bar X_{\ell m} = _{\pm 2}\bar P_{\ell m}= -\left(\bar E_{\ell m} \pm i\bar B_{\ell m} \right).
\end{equation}
In the notation of Ref.~\cite{Aghanim:2018oex}, these maps are the result of the filtering step $\bar X = \mathcal B^\dagger \Cov^{-1} X ^\textrm{\rm dat}$, where $X^{\rm dat}$ is the (beam-convolved) data, $\Cov$ its covariance matrix, and $\mathcal B$ the beam and transfer function mapping the CMB skies to the pixelized data. This inverse-variance filtering operation is trivial (diagonal in harmonic space) for perfectly isotropic data, but in a realistic setting with masked pixels, inhomogeneous noise, etc, this is difficult to perform and several schemes are available with slightly different results.

Most formulae exposed in this document can be derived through simple application of this formal relation,
\begin{widetext}
\begin{equation}
\begin{split}
&\sum_{m_1,m_2}\int d^2n_1\:_{s_1} Y_{\ell_1 m_1}(\hn_1)\:_{s_2} Y_{\ell_2 m_2}(\hn_1)\:_{r_1} Y_{L M}(\hn_1)\int d^2n_2\:_{t_1} Y_{\ell_1 m_1}(\hn_2)\:_{t_2} Y_{\ell_2 m_2}(\hn_2)\:_{r_2} Y_{L' M'}(\hn_2)  \\&= \delta_{LL'}\delta_{MM'}\frac{2\ell_1 + 1}{4\pi}\frac{2\ell_2 + 1} {4\pi} 2\pi \int_{-1}^{1} d\beta \: d^{\ell_1}_{s_1,t_1}(\beta)d^{\ell_2}_{s_2 t_2}(\beta)d^{L}_{r_1 r_2}(\beta) \quad (\textrm{whenever } s_1 + s_2 + r_1  = 0 = t_1 + t_2 + r_2).
\end{split}
\end{equation}
\end{widetext}
where $d^\ell_{mm'}$ are Wigner small d-matrices. We adopt the convention, standard in CMB lensing, to write quadratic estimator multipoles with $L, M$ and use $\ell, m$ for the CMB fields from which they are built.
\section{Gaussian covariance calculations}\label{sec:biases}
For two generic estimators as defined in Eq.~\eqref{eq:QE}, we now obtain the four gradient (g) and curl (c) variances and covariances with two one-dimensional integrals. These terms are often denoted with $N_L^{(0)}$ after proper normalization of the estimators. They act as leading noise biases for estimators of the anisotropy source gradient and curl spectra and cross-spectra.

As is obvious from Eq.~\eqref{eq:QE}, the result will combine spectra and cross-spectra of the inverse-variance filtered maps $\bar X$. Since there are distinct schemes to perform the inverse-variance filtering in a realistic case, we can make here a distinction between the semi-analytical and analytical $N_L^{(0)}$. In the former case, some empirical estimate of these spectra are used (assuming isotropy of the filtered maps). This gives a realization-dependent estimate of the noise. In the latter case, some analytic isotropic approximation of the filtering function is used, giving a realization-independent noise estimate. Neither are as accurate as the much more costly realization-dependent noise debiaser (`RDN0') now routinely used in CMB lensing analyses\cite{Aghanim:2018oex, Story:2014hni, Sherwin:2016tyf}.

For an isotropy estimator $_{r}\hat q$ let $s = (\si, \so, w^{\si\so})$ collectively describes the in and out spins and weight function of the left leg, and similarly with $t$ for the right leg (by consistency, $\so + \to = r$). In the same way, let $u$ and $v$ describes another estimator $_{r'}\hat q'$ (with $\uo + \vo = r'$). Then, their Gaussian correlation functions are
\begin{equation}\label{eq:cfnoise}
	\xi^{rr'}_{\pm}(\beta) = \xi^{\pm s, u}(\beta) \xi^{\pm t, v}(\beta) +  \xi^{\pm s, v}(\beta) \xi^{\pm t, u}(\beta),
\end{equation}

where $\xi^{s,t}$ is
\begin{equation}
\xi^{s,t}(\beta) \equiv  \sum_\ell \left(\frac{2\ell + 1}{4\pi}\right)w^{\so\si}_\ell w^{*\to\ti}_\ell \bar C_\ell^{\si \ti} d^\ell_{\so\to}(\beta)
\end{equation}
and $\bar C_\ell^{\si \ti} \equiv \av{ _{\si}\bar X_{\ell m}\: _{\ti} \bar X^*_{\ell m} }$.
Let $N_L^{(0, g^r g^{r'})}$ denote the Gaussian covariance of the gradient modes of the estimators
\begin{equation}
\delta_{LL'} \delta_{MM'} N_L^{(0, g^r g^{r'})}  \equiv \left.\av{\hat g^{r}_{LM} \hat g^{*, r'}_{L' M'} }\right|_{\rm G.}	
\end{equation}
and similarly for the curl-curl, gradient-curl and curl-gradient spectra.
Projecting the correlation functions in Eq.~\eqref{eq:cfnoise} onto gradient and curl modes results in

\begin{equation}\label{eq:N0gc}
\begin{split}
%\left.\av{\hat g^{r}_{LM} \hat g^{*, r'}_{L' M'} }\right|_{\rm G.}
N_L^{(0, g^rg^{r'})}&=  \frac 12 \Re\left[C_L^{rr'} +  (-1)^{r} C_L^{-rr'}\right] \\
		N_L^{(0, c^rc^{r'})} &=  \frac 12 \Re\left[C_L^{rr'} -  (-1)^{r} C_L^{-rr'}\right]\\
	N_L^{(0, g^rc^{r'})} &=  \frac 12 \Im\left[-C_L^{rr'} -  (-1)^{r} C_L^{-rr'}\right] \\ N_L^{(0, c^rg^{r'})} &=  \frac 12 \Im\left[C_L^{rr'} -  (-1)^{r} C_L^{-rr'}\right]
\end{split}
\end{equation}
where \begin{equation}\label{eq:N0C}
C_L^{\pm rr'}  \equiv 2\pi  \int_{-1}^1 d \mu\:  d^L_{\pm rr'}(\mu) \xi^{rr'}_{\pm}(\beta).
\end{equation}
$\Re$ and $\Im$ stands for real and imaginary parts. 
\newline
\newline
A non-zero gradient-curl mode cross-covariance $N_L^{(0, g^rc^{r'})}$ or $N_L^{(0, c^rg^{r'})}$ may be sourced by gradient-curl couplings in the inverse-variance filtered CMB fields (i.e., non-zero $C_\ell^{\bar T \bar B}$ or $C_\ell^{\bar E \bar B}$, for example from polarization angle miscalibration or other systematics). This is not the only possibility though.

\section{Response and cross-responses calculations}
We now turn to the calculation of the responses of a quadratic estimator given by Eq.~\eqref{eq:QE} to a source of anisotropy. In order to do this, we need to parametrize anisotropy, which we do in section~\ref{sec:param}.

Responses are only isotropic in a idealized situation (in the absence of masking and other complications). In this case we can write the inverse-variance filtering step  (beam-deconvolved) data maps $_sX^{\rm dat}$ with the help of a matrix $F$, diagonal in harmonic space,
\begin{equation}\label{eq:filter}
	_{s}\bar X_{\ell m} \equiv \sum_{s_2 = 0,2,-2}F_\ell^{s s_2} \:_{s_2}X^{\rm dat}_{\ell m}.
\end{equation} 
\subsection{Response parametrization}\label{sec:param}
Anisotropy can sometimes be parametrized at the level of the CMB maps, with
\begin{equation}\label{eq:mapresp}
	_{s}\delta X^{\rm cmb}(\hn) = \sum_{a = \pm r}\:_{a}\alpha(\hn) \left( \sum_{\ell m}\: R_\ell^{a, s} \:_sX^{\rm  resp}_{\ell m} \Ylm {s- a}(\hn)\right)
\end{equation}
for response kernel functions $R^{r,s}_\ell$, and response field $_sX^{\rm resp}$. In most situations the response field is the CMB itself. More generally, let the covariance of the CMB (beam-deconvolved) data respond as follows to a spin-$r$ anisotropy source $\alpha$:
\begin{widetext}
\begin{equation}\label{eq:covresp}
	\delta  \av{_sX^{\rm dat}(\hn_1) \:_tX^{*\rm dat}(\hn_2)} =   \sum_{\ell m, a = \pm r}\:_{a}\alpha(\hn_1) W_\ell^{a, st} \:_{s - a}Y_{\ell m}(\hn_1)  \:_{t}Y^*_{\ell m}(\hn_2)  +   W_\ell^{* a, ts} \:_{s}Y_{\ell m}(\hn_1)  \:_{t-a}Y^*_{\ell m}(\hn_2)\:_{a}\alpha^*(\hn_2)
\end{equation}
\end{widetext}
for some weights functions $W_\ell^{a, st}$. For map-level descriptions in Eq.~\eqref{eq:mapresp} then holds
\begin{equation}\label{eq:R2W}
	W_\ell^{a, st} = R^{a, s} C_\ell^{st}, \quad \textrm{ with } \quad  C_\ell^{st}\equiv \av{\:_sX_{\ell m}^{\rm resp}\: _tX_{\ell m}^{\dagger \rm cmb}}
\end{equation}
However, Eq.~\eqref{eq:covresp} is more general. Section~\ref{sec:examples} lists weights functions of some relevant cases.

\subsection{Estimator responses calculation}\label{sec:responses}

Let as before $s, t$ denote collectively the QE spins and weight functions for an estimator $_r\hat q(\hn)$ of spin $r = s_o + t_o$, and let $r'$ be the spin of anisotropy source $_{r'}\alpha(\hn)$ with covariance response kernel $W^{r'}$ as above. Let $\mathcal R_L^{g_r g_{r'}} \delta_{LL'}\delta_{MM'}$ be defined as the response of the gradient mode $g^{r}_{LM}$ of $_{r}\hat q$ to the gradient mode $g^{r'}_{L'M'}$ of $_{r'}\alpha$, and similarly for the curl. It holds: \begin{equation}\label{eq:RLgc}%\boxed{
	\begin{split}
		\resp^{g_rg_{r'}}_L &= \Re\left[R_L^{st, r'} + (-1)^{r'} R_L^{st, -r'}\right]\\
		\resp^{c_rc_{r'}}_L &= \Re\left[R_L^{st, r'} - (-1)^{r'} R_L^{st, -r'}\right] \\
		\resp^{g_rc_{r'}}_L &= \Im\left[-R_L^{st, r'} + (-1)^{r'} R_L^{st, -r'} \right]  \\
		\resp^{c_rg_{r'}}_L &= \Im\left[R_L^{st, r'} + (-1)^{r'} R_L^{st, -r'} \right] \\
	\end{split}
	%}
\end{equation}
where
\begin{widetext}
\begin{equation}\label{eq:RL}
%\boxed{
\begin{split}
R_L^{st, r'} &= 2\pi  \int_{-1}^1 d \mu\: d^L_{rr'}(\mu)\sum_{\tilde s_i,\tilde t_i = 0,2,-2}  \left[\xi^{\so \si \tilde s_i} (\mu)\psi^{\to \ti \tilde t_i \tilde s_i, r' }(\mu) +  \xi^{\to \ti \tilde t_i}(\mu) \psi^{\so \si \tilde s_i \tilde t_i, r' }(\mu) \right]
\end{split}
%}
\end{equation}
and
\begin{equation}\label{eq:Rxipsi}
%\boxed{
\begin{split}
\xi^{\so\si \tilde s_i}(\mu) &\equiv  \sum_\ell \left(\frac{2\ell + 1}{4\pi}\right)w^{\so\si}_\ell F_\ell^{\si \tilde \si} d^\ell_{\so,\tilde \si}(\mu) 
\\
\psi^{\so\si \tilde \si \tilde \ti, r'}(\mu) &\equiv \sum_\ell \left(\frac{2\ell + 1}{4\pi}\right)w_\ell^{\so \si}F^{\si \tilde \si}_\ell W_\ell^{*-r', -\tilde \ti \tilde \si} d^\ell_{\so,-\tilde \ti + r'}(\mu) 
\end{split}
%}
\end{equation}
	
\end{widetext}
Again, in most relevant cases (but not always), the gradient to curl and curl to gradient responses do vanish. If there is a unique source of anisotropy, properly normalized gradient and curl estimators are then given by $\hat g^r_{LM} / \mathcal R_L^{g_rg_r}$ and $\hat c^r_{LM} / \mathcal R_L^{c_r c_r}$.

\section{Derivation of optimal QE weights}\label{sec:opti}Optimal (defined in the sense of minimal Gaussian variance) QE weights can be easily gained from the representation of the anisotropy. In the presence of the anisotropy ($_{r}\alpha(\hn)$ in Eq.~\ref{eq:covresp}), the CMB remains Gaussian. For small anisotropy, a good estimate will be provided by the leading term in a Newton-Raphson estimate of $_{r}\alpha$\citep{Hanson:2009gu}. This first iteration is given by the gradient of the log-likelihood $\ln p(\alpha, X^{\rm dat})$, normalized by the log-likelihood Hessian (on average equal to the Fisher matrix), all evaluated at zero anisotropy. We can combine un-normalized estimates of the real ($\Re \:_r{\alpha}$) and imaginary ($\Im \:_r{\alpha}$) parts of $_{r}\alpha$ to a spin-$r$ un-normalized estimate
with the rule
 \begin{equation}
 \frac{\delta }{\delta \Re\: _{r}\alpha(\hn)} + i \frac{\delta }{\delta \Im\: _{r}\alpha(\hn)} =2 \frac{\delta }{\delta _{- r}\alpha (\hn)}
\end{equation}
where the functional derivative with respect to $_{-r}\alpha(\hn)$ is performed in the usual way, treating $_{-r}\alpha(\hn)$  and $_{r}\alpha(\hn)$ as independent variables. 
Written in full, this is
\begin{equation}\label{eq:optiqe}
\begin{split}
	_{r}\hat \alpha(\hn) \equiv &-\left.\frac{\delta }{\delta _{- r}\alpha (\hn)} \: _{s_1}X^{\rm dat} \Cov^{-1}_{s_1s_2} \:_{s_2}X^{\rm dat} \right|_{\alpha \equiv 0}  - \left.\frac{\delta }{\delta\: _{- r}\alpha(\hn)} \ln \det \Cov \right|_{\alpha \equiv 0} 
\end{split}
\end{equation}
where $\Cov_{s_1 s_2}(\hn_1, \hn_2) \equiv \av{_{s_1}X^{\rm dat}(\hn_1) \:_{s_2}X^{\dagger\rm dat}(\hn_2) }$ is the data covariance matrix. The second term is the likelihood determinant variation. This determinant term is equal to the average of the quadratic estimate just defined, and called for this reason the `mean-field' (see Ref.~\cite{Hanson:2009gu}). In practice this mean-field is estimated as the average of the quadratic term using a realistic set of simulations.

Performing the derivative using representation~\eqref{eq:covresp} we find (neglecting the mean-field part)

\begin{equation}
	_{r}\hat \alpha(\hn) = \sum_{s,t} \frac{1}{_s2}\: _{-s}\bar X(\hn)\cdot \left( \sum_{\ell m}2W_{\ell}^{-r, st} \:\frac{1}{_t2}\:_{t}\bar X_{\ell m}\: _{s + r}Y_{\ell m}(\hn)\right)
\end{equation}
where the symbol $_s{2}$ stands for either $1 (s = 0)$ or $2 (s \ne 0)$. Hence, confronting to definition~\eqref{eq:QE},
\begin{equation}
	w_\ell^{st} = \frac{ \delta_{st} }{_s 2}\textrm{   (1st leg)  } \quad 	w_\ell^{-s + r, t} = \frac{2}{_t 2}W^{-r, -st}_\ell \textrm{   (2nd leg)  }
\end{equation}
Restricting the $s,t $ spins to $0$ or $\pm 2$ in Eq.~\eqref{eq:optiqe} provides temperature-only or polarization-only estimators. This derivation recovers for example the minimum-variance (MV), temperature-only and polarization-only lensing estimators as obtained first by Ref.~\cite{Okamoto:2003zw}(for the gradient term), with identical normalization. Usage of the lensed spectra\citep{Hanson:2010rp} instead of the unlensed spectra (or even better, the grad-lensed spectra\citep{Lewis:2011fk, Fabbian:2019tik}) are well-known slight modifications that improve the estimators. Further restrictions by zeroing maps on the left or right legs gives additional estimators (TE, TB, EB).

\section{Examples}\label{sec:examples}
The construction of optimal estimators only requires knowledge of the data covariance response $W$, Eq.~\eqref{eq:covresp}. When this is known, Eq.~\eqref{eq:QE} provides the optimal QE estimate of both gradient and curl modes, with a small number of spin-weight harmonic transforms. In the case that the anisotropy is defined at the levels of the CMB fields rather than the data covariance, $W$ is trivially related to the field response $R$ through Eq.~\eqref{eq:R2W}. In this section, we briefly derive and list a few relevant cases of these responses.
\\ \\
\textbf{Lensing:} 
	The source of anisotropy is the spin-1 deflection field $_1\alpha(\hn)$, with linear response (see Ref.~\cite{Challinor:2002cd})
	$\delta _sX^{\rm cmb}(\hn) =  -\frac 12 \alpha_1(\hn) \bar \eth _{s}X^{\rm cmb}(\hn) - \frac 12 \alpha_{-1}(\hn) \eth \:_sX^{\rm cmb}(\hn) $
	where $\eth$ and $\bar \eth$ are the spin raising and spin lowering operator respectively. From the action of these operators on the spherical harmonics follow immediately
	\begin{equation}
	\begin{split}	
		R_\ell^{-1, s} =- \frac 12\sqrt{ (l - s) (l + s + 1) }, \quad
		R_\ell^{1, s} = +\frac12\sqrt{ (l + s) (l - s + 1) }
	\end{split}
	\end{equation}
\\ \\
\textbf{Modulation, patchy reionization:}
	When searching for a large-scale trispectrum $\tau_{NL}$ signature in the CMB\citep{Pearson:2012ba, Ade:2013ydc}, the anisotropy source is a scalar (`f'), with response $\delta _sX^{\rm cmb}(\hn) = \:f(\hn) _{s}X^{\rm cmb}(\hn)	$, hence
	$R_\ell^{0,s} = 1$. Estimators of the optical depth $\hat \tau(\hn)$ in the context of patchy reionization as defined in Ref.~\citep{Dvorkin:2008tf} are similar ($R_\ell^{0,s} = 1$). The response fields $X^{\rm resp}$ are however different from  $X^{\rm cmb}$ resulting in different cross-spectra in the covariance response function in Eq.~\eqref{eq:R2W}.\\ \\	
\textbf{Polarization rotation:} 
	  Polarization angle mis-calibration, cosmic birefringence or very tiny high order large-scale structure effects could for example rotate the polarization before it is observed. If we introduce $\beta(\hn)$ according to $_{\pm 2} P^{\rm cmb} \rightarrow e^{\mp 2i \:\beta(\hn)} \: _{\pm 2}P^{\rm cmb}(\hn)$, then $R_\ell^{0, \pm 2} = \mp 2i$.
\\ \\ 
\textbf{Point sources in temperature:} (`$S^2$', see Ref.~\cite{Osborne:2013nna}): here anisotropy is sought in the intensity field of the form
	$\delta  \av{T^{\rm cmb}(\hn) \:T^{\rm cmb}(\hn')} \ni \delta^D(\hat n-\hat n')S^2(\hn)$. Hence,
	$W^{r, st}_\ell = \frac 14\delta_{r0}\delta_{s0}\delta_{t0} $.
\\ \\
\textbf{Noise variance map inhomogeneities:}
	We can look for inhomogeneities in the pixel noise variance $\sigma^2(\hn)$ across the sky. This leaves an anisotropy signature on the diagonal of the CMB data covariance matrix similar than that of point sources, with the difference that point-sources are convolved with the instrument transfer function but the pixel noise is not. $\delta  \av{T^{\rm dat}(\hn) \:T^{\rm dat}(\hn')} \ni \delta_{\hat n\hat n'}\sigma^2(\hn)$. Hence, $W^{r, st}_\ell = \frac 14\delta_{r0}\delta_{s0}\delta_{t0}  \frac{1}{b_\ell^2}$
\vspace{1cm}

Many thanks to Antony Lewis, Anthony Challinor, as well as Duncan Hanson. Support by the European Research Council under the European Union's Seventh Framework Programme (FP/2007-2013) / ERC Grant Agreement No. [616170] is acknowledged.

\bibliography{lensingbib.bib}
\end{document}